%% file: ups_prl.tex
\documentclass[aps,prl,showpacs,twocolumn,groupedaddress]{revtex4}  
\usepackage{graphicx}  % needed for figures
\usepackage{dcolumn}   % needed for some tables
\usepackage{bm}        % for math
\usepackage{amssymb}   % for math
\begin{document}
%
% the following information is for internal review, please remove them for submission
%\leftline{Version 2.3 \today, {\bf Deadline for Comments was 25/01/05}} 
%\leftline{Primary authors:  Daniela Bauer, Jundong Huang, Andrzej Zieminski}
%\rightline{Comments to {\tt d0-run2eb-007@fnal.gov}}
%
% Does not work with preprint style
%\hspace{5.2in} \mbox{FERMILAB-PUB-05/020-E}
%
\title{Measurement of inclusive differential cross sections for $\Upsilon$(1{\sl S}) production in $p\bar{p}$
  collisions at $\sqrt{s}$ = 1.96 TeV}
\input list_of_authors_r2.tex  % input Dzero author list
\date{\today}
\begin{abstract}
We present measurements of the inclusive production cross sections of 
the $\Upsilon (1S)$ bottomonium state in $p\bar{p}$ collisions at  $\sqrt{s}$ =
1.96 TeV. Using the $\Upsilon (1S)\rightarrow \mu^+\mu^- $ decay mode
for a data sample of 159 $\pm$ 10  pb$^{-1}$ collected by the D\O ~detector at
the Fermilab Tevatron collider, we determine the differential cross sections as 
a function of the $\Upsilon (1S)$ transverse momentum for three ranges of the 
%$p_T^{\Upsilon}$
$\Upsilon (1S)$ rapidity: $0 < |y^{\Upsilon}| \leq 0.6, 0.6 < |y^{\Upsilon}| \leq 1.2 $, 
and $1.2 < |y^{\Upsilon}| \leq 1.8 $.  
\end{abstract}
%
% There didn't seem to be a fitting category --- best guess
\maketitle
%
%%%%%%%%%%%%%%%%%%%%%%%%%%%%%%%%%%%%%%%%%%%%%%%%%%%%%%%%%%%%%%%%%
%\newpage
%\section{\label{sec:intro}Physics motivation}
%%%%%%%%%%%%%%%%%%%%%%%%%%%%%%%%%%%%%%%%%%%%%%%%%%%%%%%%%%%%%%%%%
%
     Quarkonium production in hadron-hadron collisions provides insight into the nature of strong interactions.
It is a window on the boundary region between perturbative and non-perturbative QCD.
Recent advances in the understanding of quarkonium production have been
stimulated by the unexpectedly large cross sections for direct $J/\psi$ 
and $\psi (2S)$ production at large transverse momentum ($p_T$)  measured at the  Fermilab
Tevatron collider~\cite{jpsi:run1-cdf-jpsi}. 
     Bottomonium states are produced either promptly or indirectly  as a result of the
decay of a higher mass state \cite{CDF:ups2}, e.g. in a radiative decay such as $\chi_b \rightarrow \Upsilon(1S) \gamma$.
%%%%%%%%%%%%%
The only detailed studies of $\Upsilon(nS)$ production at the Tevatron have been  
done by the CDF
Collaboration \cite{CDF:ups2, CDF:ups3} in the limited $\Upsilon$ rapidity range of
$|y^{\Upsilon}| < 0.4$ at $\sqrt{s}$ = 1.8 TeV,  where $y =
\frac{1}{2}\ln\frac{E+p_z}{E-p_z}$, $E$ is the $\Upsilon$ energy, and $p_z$ is the
$\Upsilon$ momentum parallel to the beam direction.
  Three types of models have been used to 
describe prompt quarkonium formation: the
color-singlet model~\cite{jpsi:mCSM}, the color-evaporation 
model~\cite{jpsi:mCEM}  (and a follow-up soft color interaction model~\cite{jpsi:mSCI}), 
and the color-octet
model~\cite{jpsi:mCOM}.  
    These models of quarkonium formation lead to different
expectations  for the production rates and polarization of the quarkonium
states, yet many of the model parameters have to be extracted directly from
the data. 
A recent paper \cite{theory:Berger, bergerd0} successfully reproduces the shape of the 
$p_T$ distribution of  $\Upsilon$ states produced at Tevatron energies by combining separate
perturbative approaches for the low- and high-$p_T$ regions. 
The absolute cross section is not predicted by these calculations, 
which are similar to the color-evaporation model.
%The next line is intentinally left blank.

%
%
%
% The next line is left blank intentionally

%
% Question 37
     In this Letter we concentrate on the production of the 
$\Upsilon(1S)$ state. A precise measurement of the differential $\Upsilon(1S)$ cross section,
using the wide rapidity range accessible by the D\O ~detector, 
will provide valuable input to the various quarkonium production models.
By reconstructing the $\Upsilon(1S)$ through its decay  $\Upsilon(1S) \rightarrow
\mu^+ \mu^-$, we determine production cross sections of the $\Upsilon(1S)$  as a function of
its transverse momentum, in three rapidity ranges: 
$0 < |y^{\Upsilon}| \leq 0.6$, $0.6 < |y^{\Upsilon}| \leq 1.2$, and $1.2< |y^{\Upsilon}| \leq 1.8$. 
%
% -------------------------------
% detector description
% -------------------------------
% the next empty line is intentional, otherwise the next paragraph has no
% indentation

%
The D\O\ detector is described in detail elsewhere \cite{run2det}. 
Here, we briefly describe only the detector components
most relevant to this analysis. 
The D\O ~tracking system consists of a high-resolution silicon microstrip
tracker (SMT) surrounded by a central scintillating-fiber tracker (CFT) 
inside a 2 T magnetic field provided by a superconducting solenoid.
The tracking volume extends to a radius of approximately 52 cm.
Closest to the interaction region is the SMT 
with a typical strip pitch of 50--80 $\mu$m. It has a
barrel-disk hybrid structure and provides tracking and vertexing coverage 
in the pseudorapidity range $|\eta| <$ 3.0,  
where $\eta = -\ln[\tan(\theta/2)]$ and $\theta$ is the polar angle. 
The CFT consists of eight concentric cylinders of pairs of scintillating-fiber
doublets. On each cylinder, the inner doublet runs parallel to the beam axis and the 
outer doublet is mounted at a stereo angle of $\pm$3$^{\circ}$, alternating with each 
cylinder.
Located outside the superconducting coil is the uranium-liquid-argon calorimeter.
Beyond the calorimeter, the muon system consists of three layers of drift tubes,
 10~cm wide in the central region ($|\eta|<1$) and  1~cm in the forward
region ($1<|\eta|<2$). Interspersed  between the drift tubes are scintillating
counters.
% primarly used for triggering. 
Located between the innermost and the 
middle layers of drift tubes are 1.8~T iron toroid magnets.
D\O ~uses up to three levels of triggers to reduce the initial event rate 
of 1.7 MHz to an output rate of approximately 50 Hz. 
% The next empty line is intentional, using \\ disables the indentation

%
%
%=========================
% \section{\label{sec:data}Data}
%% ***************************************
\begin{figure}
\begin{center}
\includegraphics[width=0.90\columnwidth]{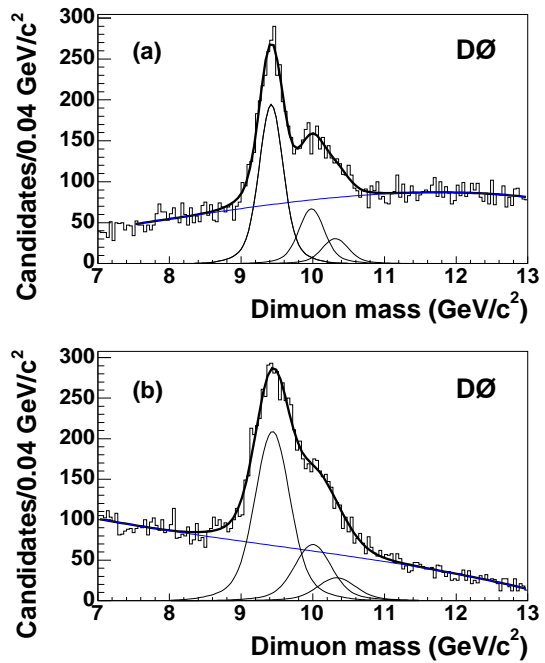}
\end{center}
\caption{Example of fits to the dimuon spectra in different bins of
  rapidity in the $p_T$ bin of 4 GeV/$c$ $< p_T^{\Upsilon} <$ 6 GeV/$c$:
(a) $|y^{\Upsilon}| \leq $ 0.6,  (b) $ 1.2 < |y^{\Upsilon}| \leq 1.8$. The heavy line shows the
  combined fit for signal and background. Also shown are the individual contributions 
  from the three $\Upsilon$ states and the background separately.}
\label{fig:mass_fits2}
\end{figure}
%%%%%%%%%%%%%%%%%%%%%%%%%%%%%
%
%    In this study, we determine production cross sections of  
%the $\Upsilon(1S)$  as a function of its
% transverse momentum, in three rapidity 
%ranges: $0 < |y^{\Upsilon}| \leq 0.6,  0.6 < |y^{\Upsilon}| \leq 1.2 $, and $1.2
%< |y^{\Upsilon}| \leq 1.8$. 
%
%We use the $\mu^+ \mu^-$ final state. 
The data were collected between June 2002 and September 2003 and  correspond to an
integrated luminosity of 159 $\pm$ 10 pb$^{-1}$  for the chosen two triggers.
These triggers are scintillator-based  dimuon triggers at the first trigger
level and  require the confirmation of one or both muons at the second trigger level. 
The first trigger level is almost fully  efficient for muons with a
transverse momentum above  5 GeV/$c$. For events passing our analysis criteria, 
the second level trigger requirement kept more 
than 97\% of events which satisfied the first level requirement. 
% We do not require a third level trigger.
% The next empty line is intentional

%
%
% eta^mu == y^mu, avoids confusion with
The analysis requires two oppositely charged muons with 
$p_T^{\mu} > 3$ GeV/$c$ and $|y^{\mu}|< 2.2$.
Only muons that are matched to a track found by the central tracking
 system and which have hits inside and outside the toroidal magnets are used.
The track  associated with a muon must have at least one hit 
in the SMT. We reject cosmic ray muons based on timing information from the muon
system scintillators.
% isolation
Compared to muons from the dominant $b\bar{b}$ background, muons 
from $\Upsilon(nS)$ decays are expected to be relatively isolated, and therefore 
we require at least one of the muons to satisfy the
following criterion:
either the sum of the transverse momenta of charged tracks in a cone of radius
0.5 (in $\eta$--$\phi$ space) around the muon is less than 1 GeV  or the sum of the 
calorimeter transverse energies in an annular cone of radii 0.1 and 0.5 around
the muon is less than 1 GeV. This isolation requirement reduces the background  
by 35\% and the signal by less than 6\%. 
% It is applied to allow a better separation of the $\Upsilon(1S)$ signal,
% however, its efficiency has to be derived from the data. 
%An additional requirement on the isolation of the $\Upsilon$ candidate itself 
%does not improve the signal to background ratio significantly and  is not
%applied.
% The next empty line is intentional

%
    Two typical examples of dimuon mass distributions in different rapidity 
bins are shown  in Fig.~\ref{fig:mass_fits2}. In each plot 
a strong $\Upsilon(1S)$ signal can be seen, accompanied  by a shoulder
attributed to unresolved 
signals due to $\Upsilon(2S)$ and $\Upsilon(3S)$ production. 
The mass distributions are fit starting from 7.0, 7.5 or 7.8 GeV/$c^2$,
depending on $p_T^{\Upsilon}$ and $y^{\Upsilon}$, to 13.0 GeV/$c^2$ using
separate mass resolution  
functions for each of the $\Upsilon (nS)$ states and a third-order polynomial for the
 background. The mass resolution function is approximated by a sum of two Gaussians with the 
relative contribution and width of the second Gaussian fixed with respect to
the first Gaussian. The values of this contribution were determined from Monte Carlo studies 
and $J/\psi$ signal fits to data. The mass of the 
$\Upsilon(1S)$ is a free parameter of the fit and the remaining two masses are shifted by the 
$m(\Upsilon(nS)) - m(\Upsilon(1S)$) differences of 563 MeV/$c^2$~$(\Upsilon(2S)$)  
and 895 MeV$/c^2$~$(\Upsilon(3S)$), taken from Ref.~\cite{PDG}. 
In addition, only the width of the $\Upsilon (1S)$ state 
is allowed to vary. 
The widths of the other states are assumed 
to scale with the mass of the resonance. Normalizations of functions representing each resonance
are free parameters of the fit. 
The Monte Carlo samples used in this study were generated with {\sc pythia} v6.202
\cite{pythia}.   
The muon kinematic distributions from data and Monte
Carlo agree within a given $p^{\Upsilon}_T$ and 
$y^{\Upsilon}$ bin.
%

%
%	
%%%%%%%%%%%%%%%%%%%%%%%%%%%%%%%%%%%%%%%%%%%%%%%%%%%%%%%%%%%%%%%%%%%%%%%%%%%%%%%%%
%\section{\label{sec:cross}Cross section calculations}
%%%%%%%%%%%%%%%%%%%%%%%%%%%%%%%%%%%%%%%%%%%%%%%%%%%%%%%%%%%%%%%%%%%%%%%%%%%%%%%%%
%
     The cross section for a given kinematic range, 
multiplied by the branching fraction $\Upsilon(1S)\rightarrow \mu^+\mu^-$, is given by:
%
%%% epsilon_kinem * epsilon_acc in conference note == epsilon_acc in PRL 
\begin{widetext}
\begin{equation}
  \frac{d^2\sigma(\Upsilon(1S))}{dp_T \cdot dy} \times
       \mathcal{B}(\Upsilon(1S) \rightarrow \mu^+ \mu^-) = \frac{N(\Upsilon(1S))}
           {{\cal L}\cdot \Delta p_T \cdot \Delta y\cdot \varepsilon_{acc}
               \cdot \varepsilon_{trig}  \cdot k_{qual} \cdot k_{trk} \cdot k_{dimu}},
\label{jpsi:cs-equation}
\end{equation}
\end{widetext}
	where ${\cal L}$ is the integrated luminosity for the data sample used, 
$N(\Upsilon(1S))$ is the number of observed $\Upsilon(1S)$, and the 
$\varepsilon$ and $k$ represent the various efficiency, acceptance and
	correction factors.
% acc
The $\Upsilon (1S)$ acceptance and reconstruction efficiency $\varepsilon_{acc}$ 
represents the fraction  of generated  $\Upsilon(1S)$ events that are successfully
reconstructed in the D\O\ detector, not taking into account any loss in
efficiency due to triggering. Its value is based  on a Monte Carlo analysis. 
% trig
The dimuon trigger efficiency $\varepsilon_{trig}$ for reconstructed
$\Upsilon(1S)$ events that satisfy our analysis criteria is estimated 
using a trigger simulation and verified directly with the data using other triggers.
The remaining factors in Eq.~\ref{jpsi:cs-equation} account for 
the differences between the data and Monte Carlo and are 
 referred to as corrections, rather than efficiencies.
% iso-smt
The  correction $k_{qual}$ takes into account differences
%between data and Monte Carlo for 
in the track quality requirements,
i.e. the isolation and SMT hit requirements and cosmic ray rejection. 
It is consistent with being independent of $p_T$ and its  
value varies between 0.85 and 0.93 with increasing rapidity.
% trk
The central tracking correction $k_{trk}$ takes into account both
differences in the tracking and the track-to-muon matching efficiency.
It is derived from the $J/\psi$ 
data sample and Monte Carlo simulation and is very close to unity except for the forward rapidity 
region where $k_{trk} \approx 0.95$. 
% dimu
The correction  factor $k_{dimu}$ accounts for the  differences in the 
local (i.e. muon system only) muon reconstruction, taking into account trigger
effects. It was determined using $J/\psi$ candidates collected with single muon triggers.
It does not show a significant $p_T$ dependence, but it changes with the muon rapidity.
%This correction factor is close to  unity for muons with $|\eta^{\mu}| < 0.3$  
%and  $|\eta^{\mu}| > 1.1$. and  drops to $ \approx 0.9$ in the intermediate region 
%between the central and forward muon detectors.
% the next empty line is intentional

%%%%%%%%%%%%%%%%%%%%%%%%%%%%%%%%%%%%%%%%%%%%%%%%%%%%%%%%%%%%%%%%%%%%%%%%%%%%%%%%%%
%================
%\section{\label{sec:results}Results}
%================
%%%%%%%%%%%%%%%%%%%%%%%%%%%%%%%%%%%%%%%%%%%%%%%%%%%%%%%%%%%%%%%%%%%%%%%%%%%%%%%%%%
% 
In Table \ref{table:cross-eff} we summarize the values of efficiencies found
in different  rapidity regions.
The measured cross sections are collected 
in Table~\ref{table:cross-total}. These cross sections are
normalized per unit of rapidity. 
\begin{table}[htp]
    \begin{center}
    \caption{Efficiencies used in the cross section calculations.}
    \begin{ruledtabular}
% PRL doesn't like vertical lines ?
%    \begin{tabular}{|c|ccccc|}
\begin{tabular}{cccccc}
%    \hline
    $|y^{\Upsilon}|$ & \scriptsize $\varepsilon_{acc}    $ & \scriptsize
      $\varepsilon_{trig}$ & \scriptsize $k_{qual}$  & \scriptsize $k_{trk}$ &
			  \scriptsize $k_{dimu}      $\\
    \hline
   0.0 -- 0.6   & 0.15 -- 0.26 &  0.70 & 0.85 & 0.99 & 0.85\\
%    \hline
   0.6 -- 1.2   & 0.19 -- 0.28 &  0.73 & 0.85 & 0.99 & 0.88\\
%    \hline
   1.2 -- 1.8   & 0.20 -- 0.27 &  0.82 & 0.93 & 0.95 & 0.95\\
%    \hline
%    \hline
    \end{tabular}
    \end{ruledtabular}
    \label{table:cross-eff}
    \end{center}
\end{table}
%%%

\begin{figure}
\begin{center}
\includegraphics[width=0.88\columnwidth]{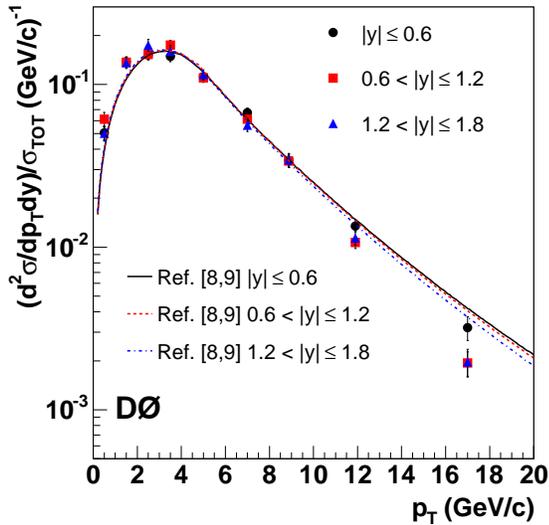}
\caption{Normalized differential cross sections for $\Upsilon (1S)$ production 
  compared
  with theory predictions \cite{theory:Berger, bergerd0}.  The errors shown
  correspond to the errors in Table \ref{table:cross-differ}.}
%    The theory curves correspond to
%  $|y^{\Upsilon}| <$ 0.6 (solid line), 0.6 $< |y^{\Upsilon}| < $ 1.8 (dashed
%  line) and 1.2 $< |y^{\Upsilon}| < 1.8$ (dotted line).}
\label{fig:theory}
\end{center}
\end{figure}

\begin{figure}
\begin{center}
%\includegraphics[width=0.99\columnwidth,
%  height=0.3\textheight]{figs/ups_ratio_prl.eps}
\includegraphics[width=0.99\columnwidth]{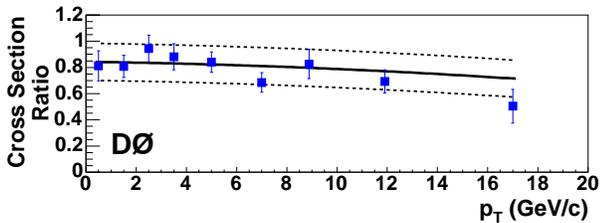}
\end{center}
\caption{
%Ratio of the $\Upsilon(1S)$ cross sections (squares) for the 
%  rapidity ranges, plotted as a function of $\Upsilon(1S)$ transverse momentum
The ratio of differential cross sections (squares) for
%$\sigma$($0.6 < |y^{\Upsilon}| \leq 1.2$) to 
 % $\sigma$($|y^{\Upsilon}| \leq 0.6$). 
$\sigma$($1.2 < |y^{\Upsilon}| \leq 1.8$) to  $\sigma$($|y^{\Upsilon}| \leq 0.6$).
The solid line is  the Monte Carlo prediction
  \cite{pythia} normalized to the measured ratio of the $p_T$-integrated cross
  section. Uncertainties of the relative normalization are indicated by the
 dashed lines.}
%\caption{Ratios of $\Upsilon(1S)$ cross sections (squares) for different
%  rapidity ranges.}
  \label{fig:cross-ratios}
\end{figure}

\begin{figure}
\begin{center}
\includegraphics[width=0.88\columnwidth]{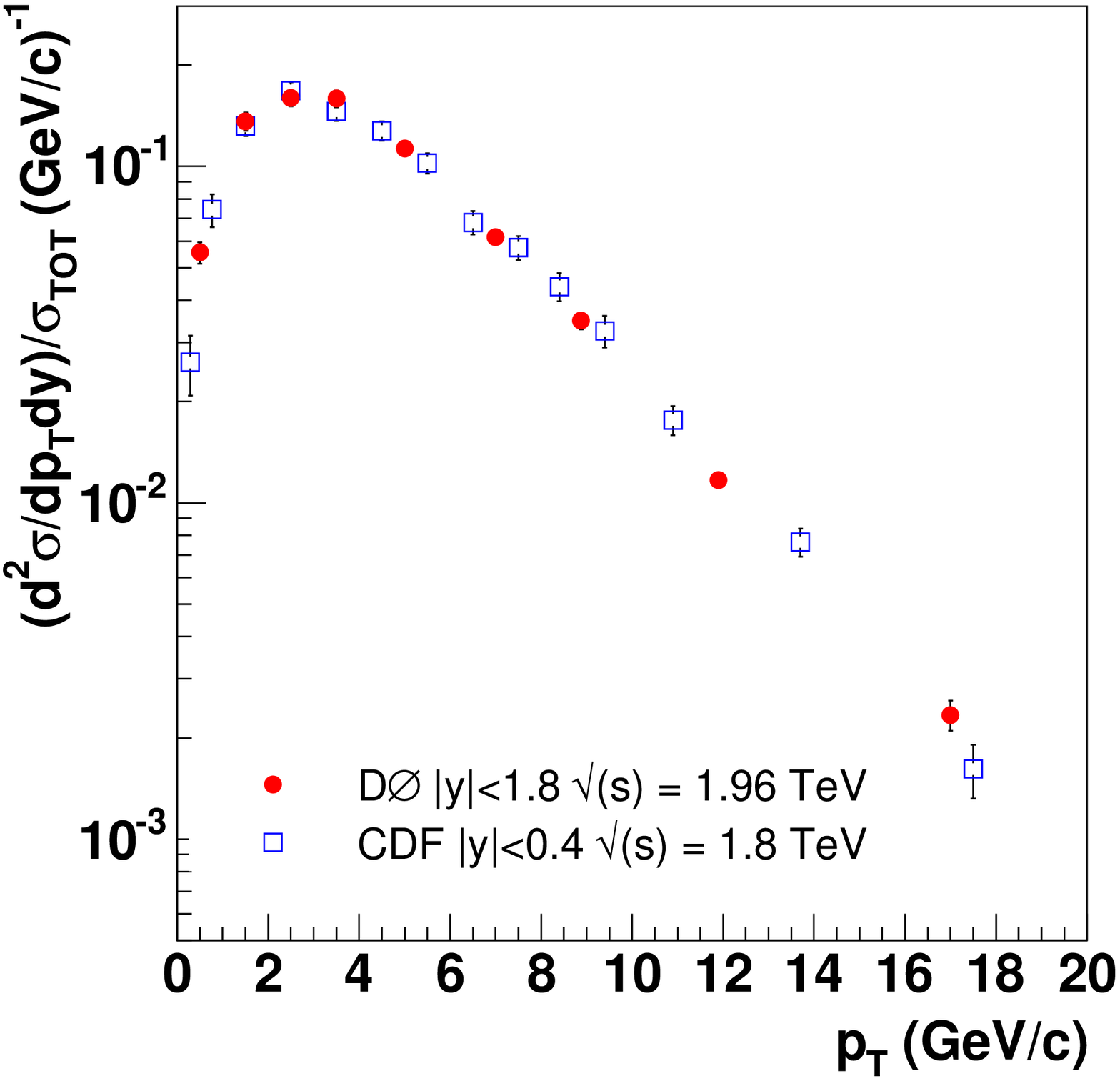}
\caption{Normalized differential cross sections for $\Upsilon (1S)$ production at $\sqrt{s}
  = 1.96$ TeV compared with published CDF results \cite{CDF:ups3} at
  $\sqrt{s} = 1.8$ TeV. The errors shown are statistical only.}
\label{fig:cross-CDF}
\end{center}
\end{figure}

%
%
% Table II
%
\begin{table*}[htp]
%    \begin{center}
\caption{Fitted number of events and 
    d$\sigma(\Upsilon(1S)$)/d$y$ $\times ~{\cal{B}}(\Upsilon(1S) \rightarrow \mu^+ \mu^-)$ per unit of rapidity.  
}    
\begin{ruledtabular}
%    \begin{tabular}{|c|cc|}
\begin{tabular}{ccc}
%    \hline
    $|y^{\Upsilon}|$ & Number of  $\Upsilon (1S)$   &
    d$\sigma(\Upsilon(1S)$)/d$y$ (pb) \\
    \hline
      0.0 -- 0.6 & 12,951 $\pm$ 336 & 732 $\pm$ 19 (stat) $\pm$ 73 (syst) $\pm$
    48 (lum) \\
%    \hline
      0.6 -- 1.2 & 16,682 $\pm$ 438 & 762 $\pm$ 20 (stat) $\pm$ 76 (syst) $\pm$
        50 (lum) \\
%    \hline
      1.2 -- 1.8 & 17,884 $\pm$ 566 & 600 $\pm$ 19 (stat) $\pm$ 56 (syst) $\pm$
        39 (lum) \\
%    \hline
      0.0 -- 1.8 & 46,625 $\pm$ 939 & 695 $\pm$ 14 (stat) $\pm$ 68 (syst) $\pm$
        45 (lum) \\
%    \hline
%    \hline
    \end{tabular}
    \end{ruledtabular}    
    \label{table:cross-total}
%    \end{center}
\end{table*}
%
%
%
%
%Table III
%==========================================================================================
 \begin{table*}[htp]
    \begin{center}
    \caption{Normalized differential cross sections for $\Upsilon (1S)$ in different rapidity regions. 
 Quoted uncertainties include statistical uncertainties added in quadrature to
 systematic uncertainties due to the assumed shape of the mass resolution
 function (cf. `stat'  uncertainties in Table \ref{table:cross-total}). 
 The remaining systematic uncertainties are $p_T$ independent and quoted in
 Table \ref{table:cross-total}.}
    \begin{ruledtabular}
    \begin{tabular}{ccccc}
% Can't get r@{$\pm$}l to work :-(
%    \hline
    $p_T^{\Upsilon}$ (GeV/$c$) & $ 0.0 <|y^{\Upsilon}| \leq 0.6 $ & $ 0.6 < |y^{\Upsilon}|
                     \leq 1.2 $ 
                     & $ 1.2 <|y^{\Upsilon}| \leq 1.8 $ & $ 0.0<|y^{\Upsilon}|
                     \leq 1.8 $\\
    \hline
      0 -- 1  & \hphantom{0}0.051 $\pm$ 0.005\hphantom{0} & \hphantom{0}0.061 $\pm$ 0.006\hphantom{0} 
              & \hphantom{0}0.050 $\pm$ 0.005\hphantom{0} & \hphantom{0}0.056 $\pm$ 0.004\hphantom{0}\\
%    \hline
      1 -- 2  & \hphantom{0}0.138 $\pm$ 0.010\hphantom{0} & \hphantom{0}0.137 $\pm$ 0.010\hphantom{0} 
              & \hphantom{0}0.136 $\pm$ 0.011\hphantom{0} & \hphantom{0}0.136 $\pm$ 0.008\hphantom{0}\\
%    \hline
      2 -- 3  & \hphantom{0}0.152 $\pm$ 0.010\hphantom{0} & \hphantom{0}0.153 $\pm$ 0.010\hphantom{0} 
              & \hphantom{0}0.175 $\pm$ 0.015\hphantom{0} & \hphantom{0}0.160 $\pm$ 0.009\hphantom{0}\\
%    \hline
      3 -- 4  & \hphantom{0}0.149 $\pm$ 0.011\hphantom{0} & \hphantom{0}0.175 $\pm$ 0.012\hphantom{0} 
              & \hphantom{0}0.160 $\pm$ 0.014\hphantom{0} & \hphantom{0}0.159 $\pm$ 0.009\hphantom{0}\\
%    \hline
      4 -- 6  & \hphantom{0}0.112 $\pm$ 0.006\hphantom{0} & \hphantom{0}0.110 $\pm$ 0.007\hphantom{0} 
              & \hphantom{0}0.115 $\pm$ 0.008\hphantom{0} & \hphantom{0}0.113 $\pm$ 0.005\hphantom{0}\\
%    \hline
      6 -- 8  & \hphantom{0}0.067 $\pm$ 0.005\hphantom{0} & \hphantom{0}0.061 $\pm$ 0.004\hphantom{0} 
              & \hphantom{0}0.056 $\pm$ 0.005\hphantom{0} & \hphantom{0}0.062 $\pm$ 0.003\hphantom{0}\\
%    \hline
      8 -- 10 & \hphantom{0}0.034 $\pm$ 0.003\hphantom{0} & \hphantom{0}0.034 $\pm$ 0.003\hphantom{0} 
              & \hphantom{0}0.034 $\pm$ 0.003\hphantom{0} & \hphantom{0}0.035 $\pm$ 0.002\hphantom{0}\\
%    \hline
     10 -- 15 & \hphantom{0}0.014 $\pm$ 0.001\hphantom{0} & \hphantom{0}0.011 $\pm$ 0.001\hphantom{0}
              & \hphantom{0}0.011 $\pm$ 0.001\hphantom{0} & \hphantom{0}0.012 $\pm$ 0.001\hphantom{0}\\
%    \hline
     15 -- 20 & 0.0032 $\pm$ 0.0005                       & 0.0019 $\pm$ 0.0003 
              & 0.0019 $\pm$ 0.0004                       & 0.0023 $\pm$ 0.0002\\
%    \hline
%     \hline
    \end{tabular}
    \end{ruledtabular}
    \label{table:cross-differ}
    \end{center}
\end{table*}
%   	                      
%
% this empty line is intentional
%
Differential cross sections, normalized to unity, are summarized in
Table~\ref{table:cross-differ}. Figure \ref{fig:theory} shows these
cross sections compared to theoretical predictions from Ref.~\cite{bergerd0}. 
There is little variation in the shape of the $p_T$ 
distributions with rapidity. This is further illustrated in
Fig.~\ref{fig:cross-ratios} which shows the ratio of the differential cross
sections of $\sigma$($1.2 < |y^{\Upsilon}| \leq 1.8$) to  $\sigma$($|y^{\Upsilon}| \leq
0.6$). 
%The slight decrease of this ratio with $p_T^{\Upsilon}$ is consistent
%with expectations from Ref. \cite{pythia}.
%
In Fig. \ref{fig:cross-CDF} we show a comparison with results from CDF \cite{CDF:ups3}.
% blank line to follow

% 
The overall systematic  uncertainties, excluding luminosity, are approximately
10\%. The uncertainty on the luminosity \cite{lumi} is 6.5\%. 
%
%%%%%%%%%%%%%%%%%%%%%%%%%%%%%%%%%%%%%%%%%%%%%%%%%%%%%%%%%%%
%================
%\section{\label{sec:syst}Systematics}
%================
%%%%%%%%%%%%%%%%%%%%%%%%%%%%%%%%%%%%%%%%%%%%%%%%%%%%%%%%%%%
% Main sources
The main systematic errors are due to the fitting procedure and the 
determination of  $k_{dimu}$.
The statistical uncertainty of the fitted number of events in a given kinematic
bin and the uncertainty from varying the contribution of the second Gaussian
are added in quadrature to give the uncertainties labeled `stat' in Table
\ref{table:cross-total}.
% and \ref{table:cross-differ}.
The net effect is an increase in the overall fit uncertainty by less 
than 40\% of its statistical uncertainty alone.
% sentence below refers to fitting range 
An additional uncertainty in the fitting procedure due to varying the fitting 
range and the background parametrization is at the 4\% level.
The systematic uncertainty for $k_{dimu}$ is 8.7\%, 8.2\% and 7.2\%
for the three rapidity bins.
%Removed, Arnd Question 20
%This systematic uncertainty reflects limited statistics and variations in the
%conditions  under which these corrections were studied. 
%Statistical and systematic uncertainties, determined as functions of kinematic
%variables describing individual muons, are added in quadrature 
%and propagated into an uncertainty for the $\Upsilon(1S)$ candidates using 
%Monte Carlo events.
These were derived from uncertainties for the Monte Carlo -- data differences
for  individual muons, determined as a function of muon rapidity and transverse momentum. 
%
%
%
%
% removed 'finite'
    The other uncertainties considered include momentum resolution, 
%variation in the hard-scatter cut off in the event generator, 
uncertainties introduced by the track quality and track matching requirements,
variations in the input Monte Carlo distributions, and changes 
in detector performance over time. All these systematic uncertainties 
contribute less than 2$\%$ each.
% the next empty line is intentional

%
    The current analysis assumes that the $\Upsilon (1S)$ is produced
unpolarized, in agreement with the CDF measurement \cite{CDF:ups3} 
of the polarization parameter $\alpha = -0.12 \pm 0.22$ for $8 < p^{\Upsilon}_T < $ 20 GeV/$c$.
Although we do not include a contribution to the systematic uncertainty due to
this assumption, we estimate the sensitivity of our results to the $\Upsilon
(1S)$ polarization by
varying $\alpha$ within $\pm 0.15$ ($\pm 0.30$). 
This changes our results by less than 4\% (15\%) in all $p_T$ bins. 
% The next empty line is intentional

%
%
%%%%%%%%%%%%%%%%%%%%%%%%%%%%%%%%%%%%%%%%%%%%%%%%%%%%%%%%%%%%%%%%%%%%%%%%%%%%%%%%%
%\section {\label{sec:conclusions}Conclusions}
%%%%%%%%%%%%%%%%%%%%%%%%%%%%%%%%%%%%%%%%%%%%%%%%%%%%%%%%%%%%%%%%%%%%%%%%%%%%%%%%%
% 'absolute' cross sections
    In conclusion, we present  a measurement of the inclusive production
cross section of the $\Upsilon (1S)$ bottomonium state using the $\Upsilon
(1S)\rightarrow \mu^+\mu^- $ decay mode. The measured cross section 
$\times$ ${\cal{B}}(\Upsilon(1S) \rightarrow \mu^+\mu^-)$ 
for the $|y^{\Upsilon}| \leq$ 0.6 region is 
732  $\pm$ 19 (stat) $\pm$ 73 (syst) $\pm$ 48 (lum)
pb. Taking into account a predicted increase in the
cross section when the $p{\bar p}$ center-of-mass energy  increases from 1.8
TeV to 1.96 TeV \cite{pythia}, our result is compatible with the CDF result
\cite{CDF:ups3} of 680 $\pm$ 15 (stat) $\pm$ 18 (syst) 
$\pm$ 26 (lum) pb for $\sqrt{s} =$ 1.8 TeV.
 We measure  the ratios of the cross sections for the 
$0.6 < |y^{\Upsilon}| \leq 1.2 $  and $1.2 < |y^{\Upsilon}| \leq 1.8 $ ranges to 
that for the $|y^{\Upsilon}| \leq 0.6 $ range to be 1.04 $\pm$ 0.14 and 0.80
$\pm$ 0.11, compared with predictions from Monte Carlo \cite{pythia} of 0.94 and
0.83. 
Between the rapidity regions, there is little variation in the shapes 
of the differential cross sections, and their shapes agree reasonably well with
theoretical predictions \cite{bergerd0}. The shape of the combined differential
 cross section for $|y^{\Upsilon}| \leq 1.8$ is consistent with the CDF
measurement  in the limited rapidity range  of $|y^{\Upsilon}| <
0.4$~\cite{CDF:ups3}. 
The results presented in this Letter will allow a more precise 
determination of parameters of the various bottomonium production models.
\begin{acknowledgments}
\input acknowledgement_paragraph_r2.tex
\end{acknowledgments}
%
%
%
%
% ***************************************
%
%
%
%
%%
% ***************************************
%
% ***************************************
%\clearpage
%
%

%
\clearpage
\end{document}

%% file: list_of_authors_r2.tex
% LIST_OF_AUTHORS_R2.TEX                 2/14/05            
%
\author{                                                                      
%% names begin here                                                           
V.M.~Abazov,$^{35}$                                                           
B.~Abbott,$^{72}$                                                             
M.~Abolins,$^{63}$                                                            
B.S.~Acharya,$^{29}$                                                          
M.~Adams,$^{50}$                                                              
T.~Adams,$^{48}$                                                              
M.~Agelou,$^{18}$                                                             
J.-L.~Agram,$^{19}$                                                           
S.H.~Ahn,$^{31}$                                                              
M.~Ahsan,$^{57}$
% ...
\\                                                                            
\vskip 0.30cm                                                                 
\centerline{(D\O\ Collaboration)}                                               
\vskip 0.30cm                                                                 
}                                                                             
\address{                                                                     
\centerline{$^{1}$Universidad de Buenos Aires, Buenos Aires, Argentina}       
\centerline{$^{2}$LAFEX, Centro Brasileiro de Pesquisas F{\'\i}sicas,         
                  Rio de Janeiro, Brazil}                                     
\centerline{$^{3}$Universidade do Estado do Rio de Janeiro,                   
                  Rio de Janeiro, Brazil}                                     
\centerline{$^{4}$Instituto de F\'{\i}sica Te\'orica, Universidade            
                  Estadual Paulista, S\~ao Paulo, Brazil}                     
\centerline{$^{5}$University of Alberta, Edmonton, Alberta, Canada,           
               Simon Fraser University, Burnaby, British Columbia, Canada,}   
\centerline{York University, Toronto, Ontario, Canada, and                    
         McGill University, Montreal, Quebec, Canada}                         
\centerline{$^{6}$Institute of High Energy Physics, Beijing,                  
                  People's Republic of China}                                 
\centerline{$^{7}$University of Science and Technology of China, Hefei,       
                  People's Republic of China}                                 
\centerline{$^{8}$Universidad de los Andes, Bogot\'{a}, Colombia}             
\centerline{$^{9}$Center for Particle Physics, Charles University,            
                  Prague, Czech Republic}                                     
\centerline{$^{10}$Czech Technical University, Prague, Czech Republic}        
\centerline{$^{11}$Institute of Physics, Academy of Sciences, Center          
                  for Particle Physics, Prague, Czech Republic}               
\centerline{$^{12}$Universidad San Francisco de Quito, Quito, Ecuador}        
\centerline{$^{13}$Laboratoire de Physique Corpusculaire, IN2P3-CNRS,         
                 Universit\'e Blaise Pascal, Clermont-Ferrand, France}        
\centerline{$^{14}$Laboratoire de Physique Subatomique et de Cosmologie,      
                  IN2P3-CNRS, Universite de Grenoble 1, Grenoble, France}     
\centerline{$^{15}$CPPM, IN2P3-CNRS, Universit\'e de la M\'editerran\'ee,     
                  Marseille, France}                                          
\centerline{$^{16}$Laboratoire de l'Acc\'el\'erateur Lin\'eaire,              
                  IN2P3-CNRS, Orsay, France}                                  
\centerline{$^{17}$LPNHE, IN2P3-CNRS, Universit\'es Paris VI and VII,         
                  Paris, France}                                              
\centerline{$^{18}$DAPNIA/Service de Physique des Particules, CEA, Saclay,    
                  France}                                                     
\centerline{$^{19}$IReS, IN2P3-CNRS, Universit\'e Louis Pasteur, Strasbourg,  
                France, and Universit\'e de Haute Alsace, Mulhouse, France}   
\centerline{$^{20}$Institut de Physique Nucl\'eaire de Lyon, IN2P3-CNRS,      
                   Universit\'e Claude Bernard, Villeurbanne, France}         
\centerline{$^{21}$III. Physikalisches Institut A, RWTH Aachen,               
                   Aachen, Germany}                                           
\centerline{$^{22}$Physikalisches Institut, Universit{\"a}t Bonn,             
                  Bonn, Germany}                                              
\centerline{$^{23}$Physikalisches Institut, Universit{\"a}t Freiburg,         
                  Freiburg, Germany}                                          
\centerline{$^{24}$Institut f{\"u}r Physik, Universit{\"a}t Mainz,            
                  Mainz, Germany}                                             
\centerline{$^{25}$Ludwig-Maximilians-Universit{\"a}t M{\"u}nchen,            
                   M{\"u}nchen, Germany}                                      
\centerline{$^{26}$Fachbereich Physik, University of Wuppertal,               
                   Wuppertal, Germany}                                        
\centerline{$^{27}$Panjab University, Chandigarh, India}                      
\centerline{$^{28}$Delhi University, Delhi, India}                            
\centerline{$^{29}$Tata Institute of Fundamental Research, Mumbai, India}     
\centerline{$^{30}$University College Dublin, Dublin, Ireland}                
\centerline{$^{31}$Korea Detector Laboratory, Korea University,               
                   Seoul, Korea}                                              
\centerline{$^{32}$CINVESTAV, Mexico City, Mexico}                            
\centerline{$^{33}$FOM-Institute NIKHEF and University of                     
                  Amsterdam/NIKHEF, Amsterdam, The Netherlands}               
\centerline{$^{34}$Radboud University Nijmegen/NIKHEF, Nijmegen, The          
                  Netherlands}                                                
\centerline{$^{35}$Joint Institute for Nuclear Research, Dubna, Russia}       
\centerline{$^{36}$Institute for Theoretical and Experimental Physics,        
                  Moscow, Russia}                                             
\centerline{$^{37}$Moscow State University, Moscow, Russia}                   
\centerline{$^{38}$Institute for High Energy Physics, Protvino, Russia}       
\centerline{$^{39}$Petersburg Nuclear Physics Institute,                      
                   St. Petersburg, Russia}                                    
\centerline{$^{40}$Lund University, Lund, Sweden, Royal Institute of          
                   Technology and Stockholm University, Stockholm,            
                   Sweden, and}                                               
\centerline{Uppsala University, Uppsala, Sweden}                              
\centerline{$^{41}$Lancaster University, Lancaster, United Kingdom}           
\centerline{$^{42}$Imperial College, London, United Kingdom}                  
\centerline{$^{43}$University of Manchester, Manchester, United Kingdom}      
\centerline{$^{44}$University of Arizona, Tucson, Arizona 85721, USA}         
\centerline{$^{45}$Lawrence Berkeley National Laboratory and University of    
                  California, Berkeley, California 94720, USA}                
\centerline{$^{46}$California State University, Fresno, California 93740, USA}
\centerline{$^{47}$University of California, Riverside, California 92521, USA}
\centerline{$^{48}$Florida State University, Tallahassee, Florida 32306, USA} 
\centerline{$^{49}$Fermi National Accelerator Laboratory, Batavia,            
                   Illinois 60510, USA}                                       
\centerline{$^{50}$University of Illinois at Chicago, Chicago,                
                   Illinois 60607, USA}                                       
\centerline{$^{51}$Northern Illinois University, DeKalb, Illinois 60115, USA} 
\centerline{$^{52}$Northwestern University, Evanston, Illinois 60208, USA}    
\centerline{$^{53}$Indiana University, Bloomington, Indiana 47405, USA}       
\centerline{$^{54}$University of Notre Dame, Notre Dame, Indiana 46556, USA}  
\centerline{$^{55}$Iowa State University, Ames, Iowa 50011, USA}              
\centerline{$^{56}$University of Kansas, Lawrence, Kansas 66045, USA}         
\centerline{$^{57}$Kansas State University, Manhattan, Kansas 66506, USA}     
\centerline{$^{58}$Louisiana Tech University, Ruston, Louisiana 71272, USA}   
\centerline{$^{59}$University of Maryland, College Park, Maryland 20742, USA} 
\centerline{$^{60}$Boston University, Boston, Massachusetts 02215, USA}       
\centerline{$^{61}$Northeastern University, Boston, Massachusetts 02115, USA} 
\centerline{$^{62}$University of Michigan, Ann Arbor, Michigan 48109, USA}    
\centerline{$^{63}$Michigan State University, East Lansing, Michigan 48824,   
                   USA}                                                       
\centerline{$^{64}$University of Mississippi, University, Mississippi 38677,  
                   USA}                                                       
\centerline{$^{65}$University of Nebraska, Lincoln, Nebraska 68588, USA}      
\centerline{$^{66}$Princeton University, Princeton, New Jersey 08544, USA}    
\centerline{$^{67}$Columbia University, New York, New York 10027, USA}        
\centerline{$^{68}$University of Rochester, Rochester, New York 14627, USA}   
\centerline{$^{69}$State University of New York, Stony Brook,                 
                   New York 11794, USA}                                       
\centerline{$^{70}$Brookhaven National Laboratory, Upton, New York 11973, USA}
\centerline{$^{71}$Langston University, Langston, Oklahoma 73050, USA}        
\centerline{$^{72}$University of Oklahoma, Norman, Oklahoma 73019, USA}       
\centerline{$^{73}$Brown University, Providence, Rhode Island 02912, USA}     
\centerline{$^{74}$University of Texas, Arlington, Texas 76019, USA}          
\centerline{$^{75}$Southern Methodist University, Dallas, Texas 75275, USA}   
\centerline{$^{76}$Rice University, Houston, Texas 77005, USA}                
\centerline{$^{77}$University of Virginia, Charlottesville, Virginia 22901,   
                   USA}                                                       
\centerline{$^{78}$University of Washington, Seattle, Washington 98195, USA}  
}                                                                             
%end                                                                          

%% file: acknowledgement_paragraph_r2.tex
% acknowledgement_paragraph_r2.tex                2/9/05
%
We thank the staffs at Fermilab and collaborating institutions, 
and acknowledge support from the 
Department of Energy and National Science Foundation (USA),  
Commissariat  \` a l'Energie Atomique and 
CNRS/Institut National de Physique Nucl\'eaire et 
de Physique des Particules (France), 
Ministry of Education and Science, Agency for Atomic 
   Energy and RF President Grants Program (Russia),
CAPES, CNPq, FAPERJ, FAPESP and FUNDUNESP (Brazil),
Departments of Atomic Energy and Science and Technology (India),
Colciencias (Colombia),
CONACyT (Mexico),
KRF (Korea),
CONICET and UBACyT (Argentina),
The Foundation for Fundamental Research on Matter (The Netherlands),
PPARC (United Kingdom),
Ministry of Education (Czech Republic),
Canada Research Chairs Program, CFI,
Natural Sciences and Engineering Research Council and 
WestGrid Project (Canada),
BMBF and DFG (Germany),
Science Foundation Ireland,
A.P.~Sloan Foundation,
Research Corporation,
Texas Advanced Research Program,
Alexander von Humboldt Foundation,
and the Marie Curie Fellowships.